\documentclass[a4paper]{article}
\usepackage{amsmath,epsfig}

\newcommand{\vb}[2]{h#1{}#2}		
\newcommand{\sct}[3]{\omega#1{}#2{}#3} 
\newcommand{\con}[3]{K#1{}#2{}#3} 
\newcommand{\nn}{\nonumber}

\begin{document}

\noindent
{\bf\Large \textsf{Analytic solutions to Riemann-squared gravity with
background isotropic torsion}}  

\vspace{0.2cm}

\noindent 
Anthony Lasenby\footnote{e-mail: \texttt{a.n.lasenby@mrao.cam.ac.uk}}, 
Chris Doran\footnote{e-mail: \texttt{c.doran@mrao.cam.ac.uk}}, and 
Reece Heineke\footnote{e-mail: \texttt{r.heineke@mrao.cam.ac.uk}}

\vspace{0.2cm}

\noindent
Astrophysics Group, Cavendish Laboratory, Madingley Road, \\
Cambridge CB3 0HE, UK.

\vspace{0.4cm}

\begin{center}
\begin{abstract}
Motivated by conventional gauge theories, we consider a theory of
gravity in which the Einstein--Hilbert action is replaced by a term
that is quadratic in the Riemann tensor.  We focus on cosmological
solutions to the field equations in flat, open and closed universes.
The gravitational action is scale invariant, so the only matter source
considered is radiation.  The theory can also accommodate isotropic
torsion and this generically removes singularities from the evolution
equations.  For general initial conditions the Hubble parameter $H(t)$
is driven in a seemingly chaotic fashion by torsion to produce
irregularly occuring inflationary regions.  In the
absence of torsion, the theory reproduces the standard cosmological
solutions of a simple big bang model.  A satisfying feature is that a
cosmological constant arises naturally as a constant of integration,
and does not have to be put into the Lagrangian by hand.
\end{abstract}
\end{center}

\section{Introduction}

Many authors have argued that gravity can be formulated as a gauge
theory~\cite{kib61,iva83,DGL98-grav}.  But there is a fundamental
difference between the Lagrangians of the electroweak and strong
interactions, and the Einstein--Hilbert action of gravity.  The
former are all quadratic in the field strength tensor, whereas the
latter is only linear.  It is therefore of interest to consider the
consequences of a theory of gravity based on a Lagrangian that is
quadratic in the Riemann tensor.  This is the theory we consider here.
We retain the gauge-theoretic framework of the
Einstein--Cartan--Kibble--Sciama theory~\cite{Grwsat}, so the tetrad
(vierbein) and connection are varied independently.  Furthermore, no
derivatives of the vierbein enter the Lagrangian, so that its
variation is straightforward and gives the quadratic analog of the
Einstein equations.

A problematic feature of the gravitational Lagrangian we consider is
that it is scale invariant.  This implies that the only matter
stress-energy tensors that can couple to gravity are trace-free.
Ultimately this symmetry would have to be broken in any physically
realistic theory.  But here we circumvent the problem by only
considering cosmological solutions where the matter is described by
radiation ($P=\rho/3$).  A more attractive property of a quadratic
Riemann Lagrangian is that torsion can be generated without an
accompanying source of spin~\cite{GADWBH}.  This is explicitly
prohibited in Ricci scalar Lagrangians \cite{Grwsat}.  The torsion is
not restricted in type, and can take the form of an isotropic field
appropriate for cosmology.  (For a review of cosmological models with
anisotropic torsion, see~\cite{Grwsat}.)

Here we construct the full cosmological field equations in a
straightforward manner, allowing for the presence of a torsion field.
The symmetries in the Lagrangian allow the equations to be expressed
in a variety of ways, and we show that one set of gauge choices leads
to a simple complex-valued differential equation.  This equations can
be solved explicitly in terms of elliptic functions, though the
accompanying evolution looks almost chaotic.  We find in particular
that the torsion couples to the Hubble parameter to produce
inflationary regimes in a quasi-periodic fashion.  While the present
day experimental upper bound on background cosmological torsion is
very small, on the order of $10^{-30}$ GeV~\cite{Shapiro,Lammer}, our
model shows that it could have been very large during some earlier
epoch, and could become large again in the future.

A further attractive feature of the theory presented here is that many
of the standard cosmological models (coupled to radiation) are
contained as special cases.  With zero torsion the equations are
solved by a standard radiation-dominated cosmology.  This set-up
illustrates neatly how the inclusion of a small amount of torsion
removes the big bang singularity by translating it into the complex
plane.  Minkevich~\cite{Minkevich} and Faria-Busto~\cite{Farina} have also
argued that a quadratic theory can remove the big bang singularity.
Furthermore, we are also able to find solutions that have a
cosmological constant present, despite the fact that is has not been
put in by hand in the original Lagrangian.  This happens because the
equations are now of higher order, and a cosmological constant arises
naturally as a constant of integration.

The paper begins with a derivation of the field equations.  Much of
the complexity of this derivation is avoided by working at the level
of the action.  We then study some of the general properties of
solutions, before considering some special cases.  Unless otherwise
stated we work in natural units where $G = c = 1$.

\section{The field equations}

Our starting point is the action integral
\begin{equation} 
S = \int d^4x \sqrt{-g}
\left(\frac{1}{12}R^{\alpha}{}_{\beta\mu\nu}R_{\alpha}{}^{\beta\mu\nu} 
+ \kappa{\cal L}_m\right) 
\end{equation}
where $R^{\alpha}{}_{\beta\mu\nu}$ is the Riemann tensor, ${\cal L}_m$
is the matter Lagrangian, and $\kappa = 8\pi$.  The gravitational
degrees of freedom in this theory are a tetrad $\vb{_\mu}{^a}$ and the
connection $\sct{_\mu}{^a}{^b}$.  Throughout we use Greek letters for
coordinate components, and Latin for tetrad components.  The Riemann
tensor is constructed from the connection by
\begin{equation}
R^{ab}{}_{\mu\nu} = \partial_\mu\sct{_\nu}{^a}{^b} -
\partial_\nu\sct{_\mu}{^a}{^b} + \sct{_\mu}{^a}{^c}\sct{_\nu}{^b}{_c}
- \sct{_\nu}{^a}{^c}\sct{_\mu}{^b}{_c},
\end{equation}
and this can be thought of as the field strength of the connection
field.

Though the theory discussed here is clearly different from general
relativity, it is still often convenient to talk in terms of
Riemannian geometry and torsion.  The metric is defined in the
expected manner by
\begin{equation}
g_{\mu\nu} = \eta_{ab} \vb{_\mu}{^a}\vb{_\nu}{^b}.
\end{equation}
The metric defines a Christoffel connection $\Gamma^\lambda_{\mu\nu}$,
and from this we can write
\begin{equation}
\sct{_\mu}{^a}{_b} = \vb{_\alpha}{^a}
(\Gamma^\lambda_{\mu\nu} + \con{^\lambda}{_\mu}{_\nu}) \vb{^\beta}{_b}
- \partial_\mu\vb{_\beta}{^a}\vb{^\beta}{_b}. 
\end{equation}
This equation defines the contorsion tensor
$\con{^\lambda}{_\mu}{_\nu}$ in terms of the pieces of
$\sct{_\mu}{^a}{_b}$ that would not be present in pure, torsion-free,
general relativity.

In this paper we are interested in cosmological solutions.  As the
underlying metric still gives rise to many of the observational
consequences of the theory, the standard cosmological arguments apply
and we can restrict to a homogeneous, isotropic line element of the
form
\begin{equation} 
(ds)^{2} = dt^{2} - a(t)^2\left(\frac{dr^2}{1 - kr^2} +
  r^2 (d\theta^2 + \sin^2\!\theta\, d\bar{\phi}^2)\right). 
\end{equation}
Here $k \in \{\pm 1,0\}$ sets the curvature, $a(t)$ is the scale
factor, and the bar on $\bar{\phi}$ is present as the $\phi$ symbol
will be used elsewhere.  A suitable tetrad for this metric is given by
\begin{equation}
\vb{_\mu}{^a} = 
\begin{pmatrix}
1 & 0 & 0 & 0 \\
0 & a(1 -  kr^2)^{-\frac{1}{2}} & 0 & 0 \\
0 & 0 & ar & 0 \\
0 & 0 & 0 &  ar\sin\theta
\end{pmatrix}.
\end{equation}

The most general spin connection $\sct{_\mu}{_a}{_b}$ consistent with
isotropy and homogeneity has components
\begin{align}
&\sct{_1}{_0}{_1} = \frac{\phi}{\sqrt{1-kr^2}},& &\sct{_1}{_2}{_3} =
  -\frac{\psi}{\sqrt{1-kr^2}},& &\sct{_2}{_0}{_2} = r\phi, \nn \\ 
&\sct{_2}{_1}{_2} = \sqrt{1-kr^2},& &\sct{_2}{_3}{_1}= -r\psi,&
  &\sct{_3}{_0}{_3} = r\phi\sin\theta, \nn \\
&\sct{_3}{_1}{_2} = -r\psi\sin\theta,& &\sct{_3}{_1}{_3} = \sqrt{1 -
  kr^2}\sin\theta,& &\sct{_3}{_2}{_3} = \cos\theta,
\end{align}
together with further terms generated by the antisymmetry in the last
two indices.  Here $\phi$ and $\psi$ are scalar functions of time.
The theory therefore contains three gravitational degrees of freedom
--- $a(t)$, $\phi(t)$ and $\psi(t)$.  It is now a straightforward
matter to compute the Riemann tensor and, in terms of a tetrad basis,
we find the following nonzero components:
\begin{align}
R^{0}{}_{j0l} &= \frac{\dot{\phi}}{a}\delta_{jl}  &
R^{0}{}_{ijk} &= -\frac{2\phi\psi}{a^2}\epsilon_{ijk} \nn \\ 
R^{i}{}_{0k0} &= -\frac{\dot{\phi}}{a}\delta^i_k &
R^{i}{}_{0jk} &= -\frac{2\phi\psi}{a^2}\epsilon_{ijk} \nn \\
R^{i}{}_{j0k} &= \frac{\dot{\psi}}{a}\epsilon_{ijk} & 
R^{i}{}_{jkl} &= \frac{(\phi^2 - \psi^2 + k)}
{a^2}\left(\delta^i_k\delta_{jl} - \delta^i_l\delta_{jk}\right)   
\end{align}
(together with further terms generated by the antisymmetry in the last
two indices).

The spatial integrals are now irrelevant to the full action integral,
and the cosmological equations are generated by the reduced action
\begin{equation}
S = \int dt  \left(a(\dot{\phi}^2 -
\dot{\psi}^2) + \frac{1}{a}(\phi^4 - 6\phi^2\psi^2 + \psi^4 + 2k\phi^2
- 2k\psi^2 + k^2) + \frac{\kappa}{3} a^3\rho\right),
\label{CosmAct}
\end{equation}
where we have assumed that the matter content is pure radiation ($p =
\rho/3$).  This implies that the matter content does not contain any
direct coupling to the spin connection.

Scale invariance is still present in this action, though in the
slightly modified form of a combined rescaling and reparameterisation.
The effect of these is that the scale factor $a$ transforms as the
inverse of an einbein,
\begin{equation} 
a(t) \mapsto a'(t) = \frac{dt}{dt'} a(t'), 
\end{equation} 
with $\phi$ and $\psi$ transforming as scalar fields, $\phi(t) \mapsto
\phi'(t) = \phi(t')$.  The result of this symmetry is that the
scale factor $a$ is essentially an arbitrary function and, after
variation, can be chosen in any convenient manner.  Two choices of $a$
are particularly attractive.  The first is the `cosmological' gauge
choice, for which
\begin{equation}
\phi = \dot{a}, 
\label{defphi}
\end{equation}
where the overdot represents differentiation with respect to $t$.
This is the gauge that most closely resembles a standard FRW
cosmology.  The evolution parameter $t$ measures cosmic time, and the
equations can be expressed in terms of the Hubble function $H(t)$,
where
\begin{equation}
 H = \frac{\dot{a}}{a}.
\end{equation}
The second gauge of interest is the `conformal' gauge in which we set
$a=1$.  In this case the evolution parameter $t$ measures conformal
time, which we denote by $\eta$.  The conformal gauge turns out to be
most convenient for solving the field equations.

\section{Equations of motion and their general solution}

The Euler--Lagrange equations for $\phi$ and $\psi$ from the action
integral~\eqref{CosmAct} are
\begin{align}
\partial_t(a\dot{\phi}) &= \frac{2}{a}(\phi^3 - 3\phi\psi^2 + k\phi) \\
-\partial_t(a\dot{\psi}) &= \frac{2}{a}(-3\phi^2\psi + \psi^3 - k\psi)
\end{align}
The $a$ variation produces a form of Hamiltonian constraint equation,
\begin{equation}
a(\dot{\phi}^2 - \dot{\psi}^2) = \frac{1}{a}(\phi^4 - 6\phi^2\psi^2 +
\psi^4 + 2k\phi^2 - 2k\psi^2 + k^2) - \kappa a^3\rho. 
\label{Ham}
\end{equation}
An immediate consequence of these equations is that we recover the
expected conservation law for radiation,
\begin{equation}
a^4 \rho = \mbox{constant}.
\end{equation}
In the absence of torsion $\psi=0$ and, working in the cosmological
gauge, the dynamics reduces to the pair of equations
\begin{align}
\ddot{H} + 4 H \dot{H} &= 2H \frac{k}{a^2} \label{torfree1} \\
(\dot{H} + H^2)^2 &= \left(H^2 + \frac{k}{a^2} \right)^2 - \kappa
\rho.
\label{torfree2}
\end{align}
We return to these equations in section~\ref{cases}.

The full equations of motions are simplest to analyse in the conformal
gauge.  Setting $a=1$, so that all derivatives are with respect to
conformal time, we arrive at the pair of equations 
\begin{align}
\phi'' &= 2\phi^3 - 6\phi\psi^2 + 2k\phi \label{phieqn} \\
\psi'' &= -6\phi^2\psi + 2\psi^3 - 2k\psi. \label{psieqn}
\end{align}
Henceforth we use dashes to denote derivatives with respect to
conformal time, and retain overdots for derivatives with respect to
cosmic time.  We now introduce the complex variable
\begin{equation}
Z = \phi + i\psi
\end{equation}
so that the equations of motion are contained in the single
complex-valued differential equation
\begin{equation}
Z'' = 2Z^3 +2kZ. 
\end{equation}
This equation has a trivial first integral, 
\begin{equation}
{Z'}^2 = Z^4 + 2kZ^2 + A_0 
\label{firstint}
\end{equation}
where $A_0$ is a complex-valued constant.  The real part of $A_0$ is
determined by the conserved energy density $a^4 \rho$ and the
curvature via
\begin{equation}
\mbox{Re}(A_0) = k^2 - \kappa a^4 \rho.
\end{equation}
The imaginary part of $A_0$ can similarly be thought of a source term
for the torsion.  The first integral~\eqref{firstint} illustrates that
the system is essentially Hamiltonian, but the complex structure
drives a mixed signature in momentum space (going as
${\phi'}^2-{\psi'}^2$), which has some unusual consequences.

\begin{figure}
\begin{center}
\includegraphics[angle=-90,width=11cm]{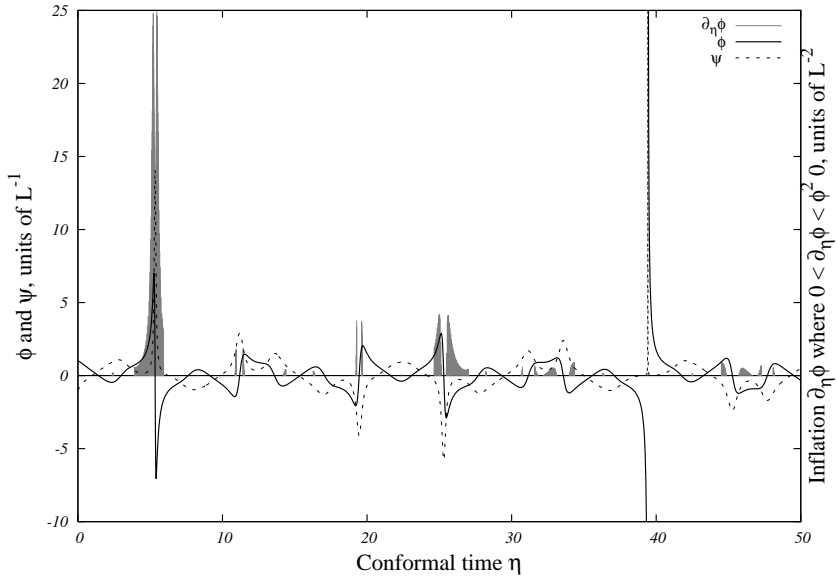}
\end{center}
\caption{\textit{Evolution of $Z$ without radiation.} The plot shows
    the real (solid) and imaginary (dashed) components of $Z(\eta)$ as
    a function of conformal time $\eta$.  For these plots we have
    $f(Z) = Z^4 - i$, $Z_0 = 1 - i$ and the sign for the initial value
    of $Z'$ is chosen so that $\phi$ is decreasing.  Also highlighted
    are the regions in which inflation is occurring, where $0< \phi' <
    \phi^2$.  The magnitude of the inflation is desribed by $\phi'$.  }
\label{Agraph}
\end{figure}

\begin{figure}
\begin{center}
\includegraphics[angle=-90,width=11cm]{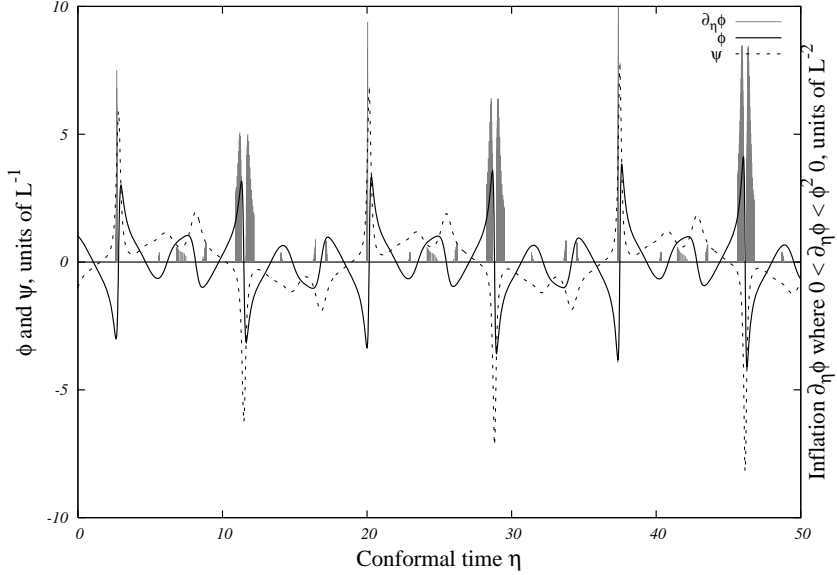}
\end{center}
\caption{\textit{Evolution of $Z$ with radiation.}  A similar graph to
    figure~\ref{Agraph}, but now with matter present.  This in
    incorporated by setting $A_0= 1 - i$.  We keep the initial
    value of $Z_0 = 1 - i$.}
\label{Bgraph}
\end{figure}

From the first integral form of the equations of motion, the full
solution can be written down as
\begin{equation}
\eta = \int \frac{dZ}{(Z^4 + 2kZ^2 + A_0)^{1/2}}. 
\label{IntSol}
\end{equation}
This integral can be solved analytically in terms of the Weierstrass
elliptic function $\wp(\eta) =
\wp(\eta;g_2,g_3)$~\cite{Whittaker,Elliptic,Weierstrass}.  This
function requires two constants, $g_2$ and $g_3$.  These are defined
in terms of the quartic polynomial
\begin{equation}
f(Z) = Z^4 + 2kZ^2 + A_0.
\end{equation}
From this we compute the pair of elliptic invariants $g_2$ and $g_3$,
obtaining
\begin{align}
g_2 &= A_0 + 12k^2 \\ 
g_3 &= 4kA_0.
\end{align}
The integral
\begin{equation} 
\eta = \int^{Z}_{Z_0} \frac{dZ'}{\sqrt{f(Z')}} 
\end{equation}
now has the general solution
\begin{equation}
Z = Z_0 + \frac{\sqrt{f(Z_0)}\wp'(\eta) +
  \frac{1}{2}f'(Z_0)\left(\wp(\eta) - \frac{1}{24}f''(Z_0)\right) +
  \frac{1}{24}f(Z_0)f'''(Z_0)}{2\left(\wp(\eta) -
    \frac{1}{24}f''(Z_0)\right)^2 - \frac{1}{48}f(Z_0)f^{(4)}(Z_0)}.  
\label{gensoln}
\end{equation}
The Weierstrass $\wp(\eta)$-function has a second order pole at $\eta
= 0$, and the derivative of the Weierstrass function, $\wp'(\eta)$, is
itself an elliptic function with pole of order 3 at $\eta = 0$.  The
constant $Z_0$ is arbitrary, and it is sometimes convenient to chose
it at one of the zeros of $f(Z)$.  With $f(Z_0) =0$ the solution
simplifies to
\begin{equation}
Z = Z_0 + \frac{1}{4}f'(Z_0)\left(\wp(\eta; g2, g3) -
  \frac{1}{24}f''(Z_0)\right)^{-1}. 
\end{equation}
Given generic initial conditions, the behavior of the general
solution~\eqref{gensoln} is extemely complicated, as the following
plots illustrate.  Figures~\ref{Agraph} and~\ref{Bgraph} show sample
behaviour both without and then with matter.  We can generally
characterize the general solution as describing periods of fairly
stable behaviour, which are quasi-periodically driven by the torsion
to produce very large, but generally finite, peaks.

The plots~\ref{Agraph} and~\ref{Bgraph} also highlight the regions
where inflation is occurring.  To define the regions of inflation we
must first recall how to switch between the conformal and cosmological
gauges.  Given $\phi(\eta)$ we can introduce a scale factor via
equation~\eqref{defphi}, the solution of which is
\begin{equation}
a(\eta) = \exp \left(\int_{\eta_0}^\eta \phi(\eta') \, d\eta' \right).
\end{equation}

\begin{figure}
\begin{center}
\includegraphics[angle=-90,width=11cm]{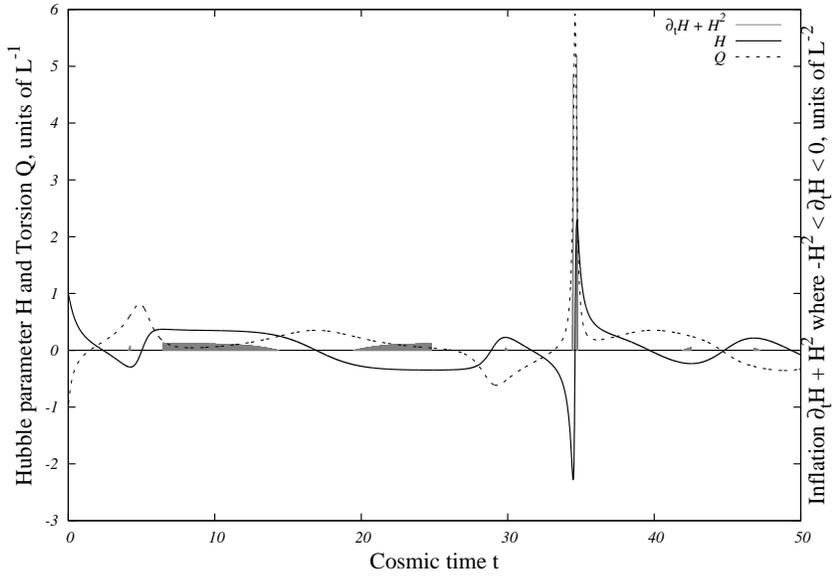}
\end{center}
\caption{\textit{Hubble function $H$ as a function of cosmic time
    $t$.} The initial conditions are chosen to match those of
    figure~\eqref{Agraph}.  Also plotted is the torsion $Q(t)$, where
    $Q=\psi/a$.  Inflationary regions are also highlighted.}  
\label{Hplot}
\end{figure}

With $a$ computed, we then compute the Hubble function by
\begin{equation}
H(\eta) = \frac{\phi(\eta)}{a(\eta)}.
\end{equation}
If necessary, $H(\eta)$ can be converted back to a function of cosmic
time, though it is easier to compute $H(t)$ from the differential
equations in the cosmic gauge.  Typical behaviour for $H(t)$ is shown
in figure~\ref{Hplot}

In the theory as it stands it makes no sense to distinguish one gauge
as being preferred over another, as we have no way of measuring $H$,
for example.  One would have to find some way of breaking the
conformal invariance and including matter with mass.  However, it is
still interesting to compute whether the solutions could represent an
inflating universe.  The standard condition for inflation is
$\ddot{a}>0$, and we also impose the further restriction that
$\dot{H}<0$.  Combined, these produce the restriction $0<-\dot{H}
<H^2$.  In terms of the conformal gauge, the equivalent condition is
\begin{equation}
0< \phi' < \phi^2.
\end{equation}
The regions for which this occurs are marked on figures~\ref{Agraph}
and~\ref{Bgraph}.

\begin{figure}
\begin{center}
\includegraphics[angle=-90,width=11cm]{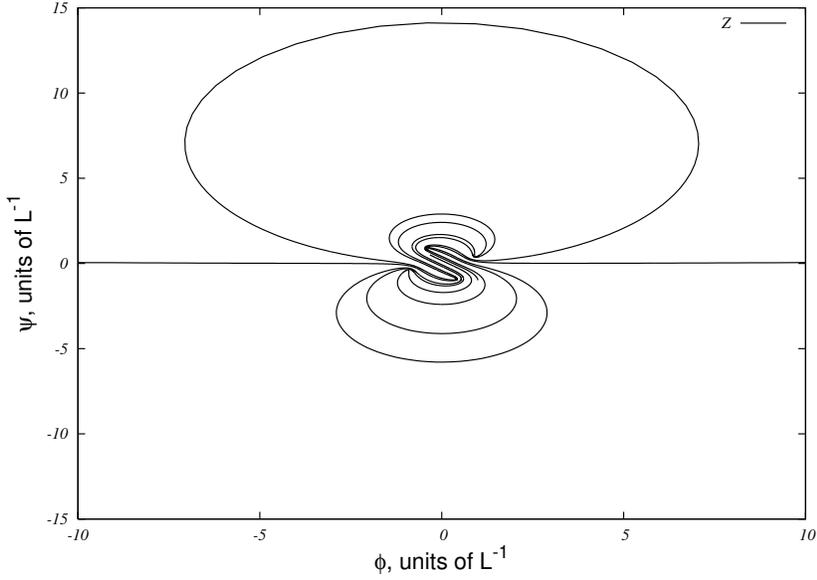}
\end{center}
\caption{\textit{Complex space trajectory of $Z$.} The plot shows the
  trajectory in complex space of $Z(\eta) = \phi + i\psi$ over a
  finite interval of conformal time.  The plot is for $f(Z) = Z^4 - i$
  with initial condition $Z_0 = 1 - i$.  The plot illustrates how the
  trajectory of $Z(\eta)$ generally stays in the vicinity of the
  origin, but occasionally moves far away.  The horiztonal line
  illustrates one such excursion.}
\label{complexspace}
\end{figure}

A further way of viewing the dynamics is shown in
figure~\ref{complexspace}.  This plot shows the trajectory of $Z(t)$
in complex space over a finite time interval.  The plot further
illustrates how the complex dynamical evolution can arise from a
simple complex ordinary differential equation.  The plot concentrates
on the fine structure around the origin and does not show the large
excursions that occur aperiodically.

\section{Special cases}
\label{cases}

In this section we consider some of the special case solutions to the
general cosmological equations.  First, returning to the torsion-free
equation~\eqref{torfree1} and integrating we find that
\begin{equation}
\dot{H} + 2H^2 = -\frac{k}{a^2} + c_0
\end{equation}
where $c_0$ is a constant of integration.  It then follows from
equation~\eqref{torfree2} that the density is given by
\begin{equation}
\frac{\kappa}{2c_0} \rho = H^2 + \frac{c_0}{2} + \frac{k}{a^2}.
\end{equation}
But, following a redefinition of $\rho$, this is precisely the
standard FRW equation with a cosmological constant $c_0/2$.  Indeed,
in order to include matter into this model, a term representing the
cosmological constant is \textit{forced}!  This is clear from the fact
that $c_0=0$ implies $\rho=0$.

If we continue to specialise by further setting $k$ and $\rho$ to
zero, the cosmological equations reduce to
\begin{equation}
\dot{H} + H^2 = \pm H^2.
\end{equation}
This has two solutions.  The positive root corresponds to constant
$H$, representing a de Sitter space.  The negative root has
$\dot{H}=-2H^2$ and is solved by $H=1/(2t)$.  This corresponds to a
simple big bang model.  So we see that many standard cosmological
solutions are contained in the quadratic Lagrangian framework.

It is instructive to see how we recover the simple big bang solution
$H=1/(2t)$ from the general solution~\eqref{gensoln}.  In the
conformal gauge, this solution is characterised by
\begin{equation}
 Z = \frac{1}{2\eta}.
\end{equation}
To recover this solution we first set $f(Z) = Z^4$, which removes the
matter and curvature contributions.  The general solution now reduces
to (\ref{gensoln}) to
\begin{equation}
Z = Z_0 + \frac{Z_0^2\wp'(\eta) + 2Z_0^3\wp(\eta)}{2\wp^2(\eta) -
  2Z_0^2\wp(\eta)}. 
\end{equation}
(Selecting the opposite sign for the square root in this expression
only changes the overall sign in the result.)  With $g_2 = g_3 = 0$ we
have
\begin{equation}
\wp(\eta) = \eta^{-2},
\end{equation}
and it follows that
\begin{equation} 
Z = \frac{1}{\eta + Z_0^{-1}}.
\end{equation}
By choosing $Z_0 \mapsto \infty$  we recover the big bang model.
An interesting effect occurs if we include some initial torsion by
setting by setting $Z_0 = -i/b$.  Now the general
solution becomes
\begin{equation} 
 Z = \frac{\pm \eta - ib}{(\eta^2 + b^2)}. 
\end{equation}
We see now that the initial big bang singularity has been removed by
the torsion and the evolution is entirely finite.  In effect, the
torsion term has moved the singularity to a complex value of the
conformal time.

\begin{figure}
\begin{center}
\includegraphics[angle=-90,width=11cm]{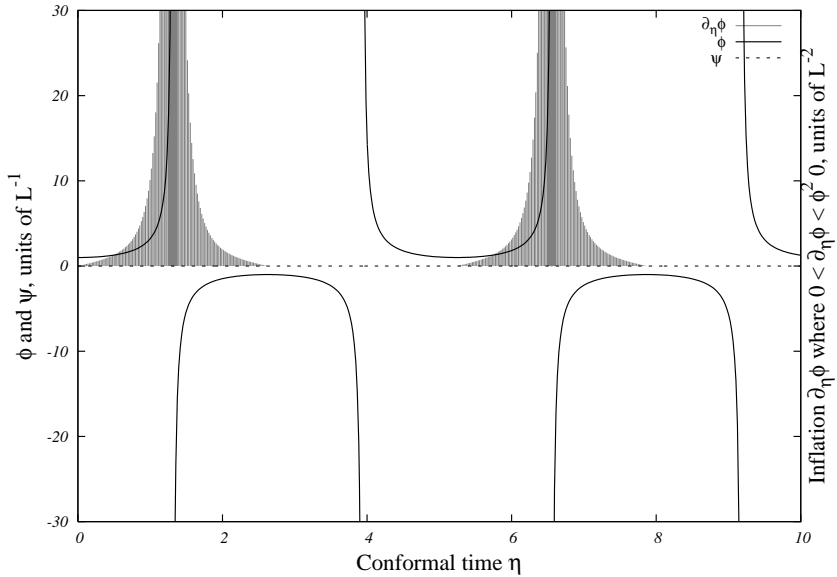}
\end{center}
\caption{\textit{Repeated big bang and big crunch scenario without
    torsion.}  Initial values are $A_0= - 1$ and $Z_0 = 1$,
    corresponding to elliptic invariants of $g_2 = -1$ and $g_3 = 0$.
    This is known as the pseudolemniscate case.}
\label{BangCrunch}
\end{figure}

We can find further interesting models within the general solution.
Choosing $f(Z) = Z^4 + 1$ and $Z_0 = 1$ then the solution produces big
bang and big crunch singularities in a periodic manner (see figure
\ref{BangCrunch}). For this configuration, $g_2 = 1$ and $g_1 = 0$.
This corresponds to one of three special cases of the elliptic
invariants that has been given a special name, this one being the
pseudolemniscate case \cite{Abram}. The pseudolemniscate case has
half-periods $L(\pm 1 + i)/4$ where $L$ is the lemniscate constant
$L/4 = 1.311028\ldots$.

\section{Conclusions}

We have shown that for a Lagrangian that is quadratic in the Riemann
tensor the cosmological equations can be solved analytically in terms
of the elliptic Weierstrass function $\wp(\eta)$.  The model can only
couple to scale-invariant fields, and the only matter considered is
isotropic radiation.  Furthermore, we have shown that these solutions
reduce to standard cosmological solutions in various limits.  This
model is particularly interesting in that it produces many periods of
inflation for a wide range of initial conditions.  But in order for
such a model to be physical it is necessary to break conformal
invariance and introduce mass, which will form the subject of future
research.


\begin{thebibliography}{10}

\bibitem{kib61}
T.W.B. Kibble.
\newblock Lorentz invariance and the gravitational field.
\newblock {\em J. Math. Phys.}, {\bf 2}(3):212, 1961.

\bibitem{iva83}
D.~Ivanenko and G.~Sardanashvily.
\newblock The gauge treatment of gravity.
\newblock {\em Phys. Rep.}, {\bf 94}(1):1, 1983.

\bibitem{DGL98-grav}
A.N. Lasenby, C.J.L. Doran, and S.F. Gull.
\newblock Gravity, gauge theories and geometric algebra.
\newblock {\em Phil. Trans. R. Soc. Lond. A}, {\bf 356}:487--582, 1998.

\bibitem{Grwsat}
F.W. Hehl, P.~von~der Heyde, G.D. Kerlick, and J.M. Nester.
\newblock General relativity with spin and torsion: Foundations and prospects.
\newblock {\em Rev. Mod. Phys.}, {\bf 48}(3):393, 1976.

\bibitem{GADWBH}
A.N.~Lasenby and C.J.L.~Doran.
\newblock Geometric algebra, Dirac wavefunctions and black holes.
\newblock In P.~G. Bergmann and V.~de~Sabbata, editors, {\em Advances in the
  Interplay between Quantum and Gravity Physics}, pages 251 -- 283. Kluwer,
  2002.

\bibitem{Shapiro}
I.L. Shapiro.
\newblock Physical aspects of the space-time torsion.
\newblock {\em Phys. Reps.}, {\bf 357}:113, 2002.

\bibitem{Lammer}
Claus L\"{a}mmerzahl.
\newblock Constraints on space-time torsion from Hughes--Drever experiments.
\newblock {\em Phys. Lett. A.}, {\bf 228}:223, 1997.

\bibitem{Minkevich}
A.~V. Minkevich.
\newblock Generalised cosmological {F}riedmann equations without gravitiation
  singularity.
\newblock {\em Phys. Lett.}, {\bf 80A}(4):232, 1980.

\bibitem{Farina}
L. Faria-Busto.
\newblock Some new cosmological results of quadratic {L}agrangians.
\newblock {\em Phys. Rev. D}, {\bf 38}(6):1741, 1988.

\bibitem{Whittaker}
E.~T. Whittaker and G.~N. Watson.
\newblock {\em A Course in Modern Analysis}.
\newblock Cambridge University Press, fourth edition, 1990.

\bibitem{Elliptic}
E.~W. Weisstein.
\newblock The Elliptic Integral, from Mathworld web resource.
\newblock http://mathworld.wolfram.com/EllipticIntegral.html.

\bibitem{Weierstrass}
E.~W. Weisstein.
\newblock The Weierstrass Elliptic Function,  from MathWorld. 
\newblock http://mathworld.wolfram.com/WeierstrassEllipticFunction.html.

\bibitem{Abram}
M.~Abramowitz and I.~A. Stegun, editors.
\newblock {\em Handbook of Mathematical Functions with Formulas, Graphs, and
  Mathematical Tables, 9th printing}.
\newblock New York: Dover, New York, 1972.

\end{thebibliography}
\end{document}